\documentclass[amsmath,amssymb,12pt]{article}
\usepackage[all]{xy}
\usepackage[centertags]{amsmath}
\usepackage{latexsym}
\usepackage{amsfonts}
\usepackage{amssymb}
\usepackage{amsthm}
\usepackage{graphicx}
\usepackage{dcolumn}
\usepackage{bm}

\def\R{{\mathbb R}}

\begin{document}

\title{Non-singular cloaks allow mimesis}

\author{Andr\'e Diatta $^{1*}$ and S\'ebastien Guenneau $^{2}$}


\maketitle
\date{}
\noindent
{\footnotesize
\thanks{\centerline{ $^{1*}$University of Liverpool. Department of Mathematical Sciences,}
\\
\centerline{ M.O. Building, Peach Street, Liverpool L69 3BX, UK}
\\
 \centerline{Email address: adiatta@liv.ac.uk;}
\\
\centerline{$^{2}$Institut Fresnel-CNRS (UMR 6133), University of Aix-Marseille,}
\newline\noindent
 \centerline{case 162, F13397 Marseille Cedex 20, France.}\\
 \centerline{Email address: sebastien.guenneau@fresnel.fr; guenneau@liv.ac.uk} }}

\begin{abstract}
We design non-singular cloaks enabling objects to scatter waves like objects
with smaller size and very different shapes.
We consider the Schr\"odinger
equation which is valid e.g. in the contexts of geometrical and quantum
optics. More precisely, we introduce a generalized non-singular
transformation for star domains, and numerically demonstrate that
an object of nearly any given shape surrounded by a given cloak scatters waves in
exactly the same way as a smaller object of another shape. When a source is located
inside the cloak, it scatters waves as if it would be located some distance away from
a small object. Moreover, the invisibility region actually hosts almost-trapped eigenstates.
Mimetism is numerically shown to break down for the quantified energies associated with
confined modes.
If we further allow for non-isomorphic transformations, our approach leads to the
design of quantum super-scatterers: a small size object
surrounded by a quantum cloak described by a negative anisotropic heterogeneous effective mass
and a negative spatially varying potential scatters matter waves like a larger nano-object of different shape.
Potential applications might be for instance in quantum dots probing. The results within this paper as well as the corresponding derived constitutive tensors,
are valid for cloaks with any arbitrary star shaped boundaries cross sections,
 although for numerical simulations, we use examples with piecewise
linear or elliptic boundaries.
\end{abstract}


\maketitle 

\section{Introduction}
Control of electromagnetic waves can be
achieved through coordinate transformations which bring exotic
material parameters \cite{pendrycloak,ulf2006a,ulf2006b}. Electromagnetic
metamaterials within which negative refraction and focussing effects
involving the near field can occur
\cite{veselago,pendry_prl00,smith00,sar_rpp05} can be understood in
light of transformation optics \cite{ulf2006b}.

\noindent Recently, an electron focussing effect across a p-n junction in a
Graphene film, that mimics the Pendry-Veselago lens in optics has
been proposed \cite{cheianov}. The subsequent theoretical
demonstration of $100$ per cent transmission of cold rubidium atom
through an array of sub de Broglie wavelength slits, brings the
original continuous wave phenomenon in contact with the quantum
world \cite{moreno}.

\noindent Other types of waves such as water waves can be controlled in a
similar way using transformation acoustics \cite{milton}, leading to
invisibility cloaks for pressure waves thanks to the design of
two-dimensional \cite{cummernjp,sanchez} and three-dimensional
cloaks \cite{pendryprl,chen07}. It has been further demonstrated
that broadband cloaking of surface water waves can be achieved with
a structured cloak \cite{farhat_prl}. Interestingly, cloaking can be further extended
to in-plane elastic waves \cite{milton,apl2009} and bending waves in thin-plates \cite{prb2009}.

\noindent In this paper, we focus our analysis on cloaking of quantum waves
which involves spatially varying potentials and anisotropic
effective mass of particles, as first proposed by the team of Zhang
\cite{matter_prl2008} and further mathematically studied by Greenleaf
et al. \cite{greenleaf-trapped}. We build up on the former proposal to render a quantum
object smaller, larger, or even change its shape. Our point here is to apply
the versatile tool of transformation physics in an area where the size of the
object might have some dramatic changes in the physics: for instance, a quantum
super-scatterer might enhance the interactions of quantum dots with the mesoscopic scale,
thereby enabling quantum effects in metamaterials.

\section{Transformed governing equations for matter waves}
Following the proposal by Zhang et al. \cite{matter_prl2008},
we consider electrons in a crystal with slowly varying composition:
$V=E_b+U$ is the spatially varying potential, $E_b$ the energy of
the local band edge and $U$ a slowly varying external potential. In
cylindrical coordinates with $z$ invariance, and letting the mass
density $m_0$ be isotropic diagonal in these coordinates, the time
independent Schr\"odinger equation takes the form
\begin{equation}
-\frac{\hbar}{2}\nabla\cdot \left(m_0^{-1} \nabla \Psi\right) + V
\Psi = E \Psi \; . \label{govpressure}
\end{equation}
Here, $\hbar$ is the Plank constant and $\Psi$ is the wave function.
Importantly, this equation is supplied with Neumann boundary
conditions on the boundary of the object to be cloaked.

\noindent Let us consider a map from a co-ordinate system
$\{u,v,w\}$ to the co-ordinate system $\{x,y,z\}$ given by the
transformation characterized by $x(u,v,w)$, $y(u,v,w)$ and
$z(u,v,w)$. This change of co-ordinates is characterized by the
transformation of the differentials through the Jacobian:

\begin{equation}
\left(%
\begin{array}{c}
  dx \\
  dy \\
  dz \\
\end{array}%
\right) = \mathbf{J}_{xu}
\left(%
\begin{array}{c}
  du \\
  dv \\
  dw \\
\end{array}%
\right) \; , \hbox{with} \; \mathbf{J}_{xu}=
\frac{\partial(x,y,z)}{\partial(u,v,w)} \;.
\end{equation}

\noindent On a geometric point of view, the matrix $\mathbf{T} \!=
\! \mathbf{J}^T \mathbf{J}/\det(\mathbf{J})$ is a representation of
the metric tensor. The only thing to do in the transformed
coordinates is to replace the effective mass (homogeneous and
isotropic) and potential by equivalent ones. The effective mass
becomes heterogeneous and anisotropic, while the potential gets a
new expression. Their properties are given by \cite{matter_prl2008}
\begin{equation}
\underline{\underline{m'}} =m_0 \mathbf{T}_T^{-1} \; , \; V'=E+
{T}_{zz}^{-1} (V-E) \; ,
\label{epsmuT}
\end{equation}
where $\mathbf{T}_T^{-1}$ stands for the upper diagonal part of the
inverse of $\mathbf{T}$ and $T_{zz}$ is the third diagonal entry of
${\bf T}$.

\noindent The transformed equation associated with the quantum mechanical
scattering problem (\ref{govpressure}) reads
\begin{equation}
-\frac{\hbar}{2}\nabla\cdot \underline{\underline{m'}}^{-1} \nabla
\Psi + V'\Psi = E \Psi \; , \label{transfpressure}
\end{equation}
where importantly the energy $E$ remains unchanged and the wave function
$\Psi(x,y)=\exp(i\sqrt{E}(xk_1+yk_2))+\Psi_d(x,y)$ with $\sqrt{k_1^2+k_2^2}=1$.
We note that $\Psi_d$ satisfies the usual Sommerfeld radiation condition
(also known as outgoing wave condition in the context of electromagnetic and acoustic
waves) which ensures the existence and uniqueness of the solution to
(\ref{transfpressure}).

It is indeed the potential $V'$ and the mass density tensor $\underline{\underline{m'}}$
(e.g. involving ultracold atoms trapped in an optical lattice as proposed in \cite{matter_prl2008})
which play the role of the quantum cloak at a given energy $E$. However, there is a
simple correspondence between the Schr\" odinger equation and the Helmholtz
equation, the energy $E$ of the former being related to the wave frequency $\omega$
of the latter via $\omega=\sqrt{E}$ (up to the normalization $c=\sqrt{2}/\sqrt{\hbar}$,
with $c$ the wavespeed in the background medium, say vacuum).
The present analysis thus covers cloaking
of acoustic and electromagnetic waves governed by a Hemholtz equation. Correspondences
bridging the current analysis with a model of transverse electric waves in cylindrical
metamatrials
are $\underline{\underline{m'}}\longleftrightarrow\underline{\underline{\varepsilon'}}$
on the one hand and $\sqrt{V'-E}\longleftrightarrow\omega$ on the other hand.

\section{Mathematical setup: Generalized cloaks for star domains}\label{chap:transformation}
This section is dedicated to a mathematical model generalizing the blowup of a point
to a transformation sending a domain to another, thus making the latter inherit the same
electronic, electromagnetic or acoustic behavior as the former, depending upon the physical context.
Although in this paper we will restrict ourselves to cloaking
regions in the plane or $3D$ Euclidean space, the transformation we propose can be readily extended to
any star domain in $\mathbb R^n,$ that is, domains  with a vantage point from which all points are
within line-of-sight. In particular, the transformation still preserves all lines passing through
that chosen fixed point.

\noindent Here is a description of the transformation in layman terms, but this could be formalised
mutatis mutandis in very abstract mathematical settings by working directly with the divergence-form
PDE of electrostatics \cite{kohn} or as the Laplace-Beltrami equation of an associated Riemannian metric
\cite{greenleaf1}. For simplicity, let us consider bounded star domains
   $D_j$ in $\mathbb R^n, ~ n=2,3$ with piecewise smooth arbitrary boundaries  $\partial D_j,$ all sharing the
    same chosen vantage point, $j=0,1,2.$ We suppose $D_2$ contains $D_1$ which in turn, contains $D_0.$
     Typically, $D_1$ is the domain to be made to mimic $D_0.$  The new transformation will be the identity
      outside $D_2$, that is, in $\mathbb R^n\setminus D_2$, but will send the hollow region
      $D_2\setminus D_0$  to the hollow region $D_2\setminus D_1$, in such a way that the boundary
       $\partial D_2$ of $ D_2$ will stay point-wise fixed, while that of $D_0$ will be mapped to
       $\partial D_1.$
  The hollow region $D_2\setminus D_1$ is meant to be the model for the cloak, endowed with Neumann conditions on its inner boundary $\partial D_1$, and in which any type of defect could be concealed, but will still have the same electronic (resp. electromagnetic or acoustic) response as the
region $D_0$ with a potential wall (resp. infinitely conducting boundary or rigid obstacle) of boundary $\partial D_0.$ In practise, we may divide the domains $D_j$ into subdomains, the part of whose boundaries lying inside $\partial D_j$ is a smooth arbitrary hypersurface.

However, in such an ideal cloaking, there is a dichotomy
between generic values of the energy $E$ (resp. wave frequency $\omega$), for which the
wave function must vanish within the cloaked region $D_1\setminus D_0$, and the discrete set of Neumann eigenvalues
of $D_1\setminus D_0$, for which there exist trapped states: waves which are zero outside of $D_1\setminus D_0$ and equal
to a Neumann eigenfunction within $D_1\setminus D_0$. Such trapped modes have been discussed in \cite{greenleaf1} when
$D_0$ vanishes.

\begin{figure}[h]
\centerline{\scalebox{0.4}{
\includegraphics[width=21cm,angle=0]{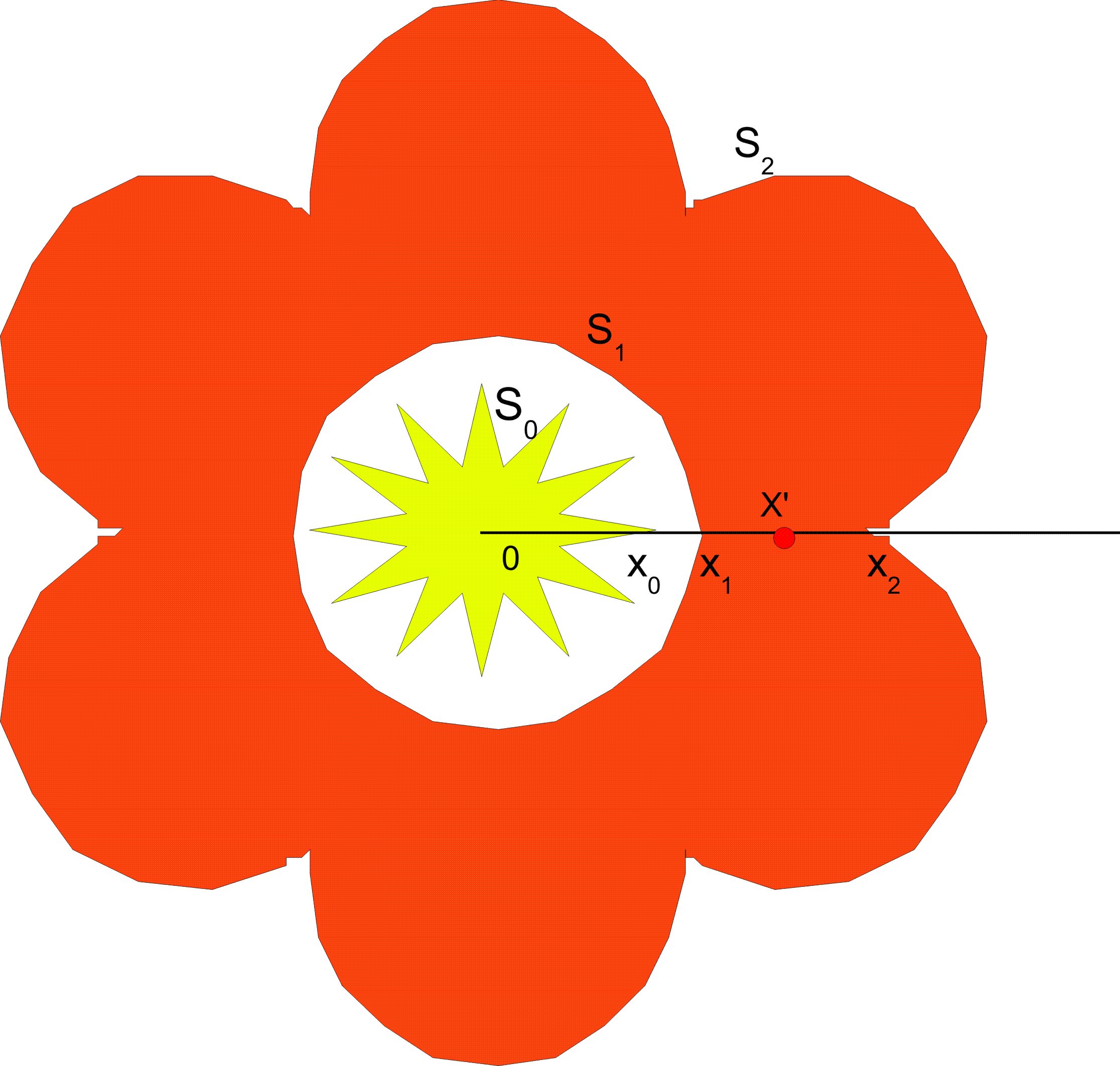}}}
\mbox{}\vspace{-0.6cm}\caption{\em \small~ Construction of a generalized non-singular cloak for mimetism.
The transformation  with inverse (\ref{eq:cylinders}) shrinks the
region bounded by the two surfaces $S_0$ and
$S_2$ into the region bounded by $S_1$
and $S_2$.
The curvilinear metric inside the carpet (here, an orange flower) is
described by the transformation matrix ${\bf T},$ see (\ref{eq:tensorgeneral})-(\ref{eq:partialderivyy}).
This is designed to play the double role of mimesis and cloaking: any types of quantum objects located within the region $D_1$ bounded by the surface $S_1$ will be invisible to an outer observer while the region itself still scatters matter waves like
an object $D_0$ bounded by $S_0$ (here, a yellow star). In the limit of vanishing yellow region, the transformation matrix ${\bf T}$
becomes singular on $S_1$ (ordinary invisibility cloak).}\label{fig:construct}
\end{figure}

\noindent The transformation is constructed as follows. Consider a point ${\bf \underline x}$ of $D_1\setminus D_0$ with ${\bf \underline x}=(x^1, x^2,...)$ relative to a system of coordinates centered at the chosen  vantage point {\bf \underline 0}. The line passing through {\bf \underline x} and  {\bf \underline 0} meets the boundaries  $\partial D_0,$ $\partial D_1,$ $\partial D_2$ at the unique points {\bf \underline x}$_0$, {\bf \underline x}$_1$, {\bf \underline x}$_2$,  respectively. We actually need the inverse ${\bf \underline x'} \mapsto {\bf \underline x}$  of the transformation, in the coordinates system, it reads $x^i=x_0^i+\alpha_i~ (x'^i-x^i_1), ~ \text{ where } ~\alpha_i=\frac{x_2^i-x^i_0}{x_2^i-x^i_1}.$ In the 3-space with coordinates $(x^1, x^2,x^3)=(x,y,z)$ we can write this transformation as
\begin{eqnarray}
\left\{
\begin{array}{lr}
x=x_0+\alpha~ (x'-x_1), ~ \text{ with } ~\alpha=\frac{x_2-x_0}{x_2-x_1}
\\
y=y_0+\beta~(y'-y_1) ~ \text{ with }~ \beta=\frac{y_2-y_0}{y_2-y_1}
\\
z=z_0+\gamma~ (z'-z_1), ~ \text{ with } ~\gamma=\frac{z_2-z_0}{z_2-z_1}
\end{array}
\right.
\end{eqnarray}

\noindent The cases of interest in this paper can all be considered as cylinders over some plane curves (triangular, square, elliptic, sun flower-like cylinders, etc.) Thus, we consider the transformation mapping the region enclosed between the cylinders ${\bf S}_0$ and  ${\bf S}_2$ into the space between  ${\bf S}_1$ and  ${\bf S}_2$ as in Figure \ref{fig:construct}, whose inverse is
 \begin{eqnarray}
\left\{
\begin{array}{lr}
x=x_0+\alpha~ (x'-x_1), ~ \text{ with } ~\alpha=\frac{x_2-x_0}{x_2-x_1}
\\
y=y_0+\beta~(y'-y_1) ~ \text{ with }~ \beta=\frac{y_2-y_0}{y_2-y_1}
\\
z=z'
\end{array}
\right.\label{eq:cylinders}
\end{eqnarray}


The matrix representation of the tensor ${\bf T}^{-1}$ is thus given by
\begin{eqnarray}
{\bf T}^{-1}=\begin{pmatrix}T^{-1}_{11}&T^{-1}_{12}&0\\
T^{-1}_{12}&T^{-1}_{22} &0\\
0&0&T^{-1}_{33}\end{pmatrix} \label{eq:tensorgeneral}\end{eqnarray}
with
\begin{eqnarray} T^{-1}_{11}=\frac{a_{11}}{a_{33}}, ~~T^{-1}_{12}=\frac{a_{12}}{a_{33}}, ~ ~ T^{-1}_{22}=\frac{a_{22}}{a_{33}},~~
T^{-1}_{33}=a_{33},\label{eq:coefgeneral1}\end{eqnarray}

\noindent where the coefficients $a_{ij}$ can be expressed as
\begin{eqnarray}
a_{11}&=&\left( \frac{\partial x}{\partial y'}\right)^2 + \left(\frac{\partial y}{\partial y'}\right)^2, ~ ~ a_{12}=-\left(\frac{\partial x}{\partial x'}\frac{\partial x}{\partial y'}+\frac{\partial y}{\partial x'}\frac{\partial y}{\partial y'}\right), \\
a_{22}&=&\left(\frac{\partial x}{\partial x'}\right)^2+\left(\frac{\partial y}{\partial x'}\right)^2, ~ ~ a_{33}=\frac{\partial x}{\partial x'}\frac{\partial y}{\partial y'}-\frac{\partial x}{\partial y'}\frac{\partial y}{\partial x'}\label{eq:coefgeneral2}\end{eqnarray}
and finally the partial derivatives are as follows
\begin{eqnarray}
\frac{\partial x}{\partial x'} &=& \frac{x_2-x_0}{x_2-x_1}+\frac{x_2-x'}{x_2-x_1}\frac{\partial x_0}{\partial x'}-\frac{(x_2-x_0)(x_2-x')}{(x_2-x_1)^2}\frac{\partial x_1}{\partial x'} \nonumber \\
&&-\frac{(x_1-x_0)(x'-x_1)}{(x_2-x_1)^2}\frac{\partial x_2}{\partial x'},\label{eq:partialderivxx}\\
\frac{\partial x}{\partial y'} &=& \frac{x_2-x'}{x_2-x_1}\frac{\partial x_0}{\partial y'}-\frac{(x_2-x_0)(x_2-x')}{(x_2-x_1)^2}\frac{\partial x_1}{\partial y'} \nonumber \\
&& -\frac{(x_1-x_0)(x'-x_1)}{(x_2-x_1)^2}\frac{\partial x_2}{\partial y'},\label{eq:partialderivxy}\\
\frac{\partial y}{\partial x'} &=& \frac{y_2-y'}{y_2-y_1}\frac{\partial y_0}{\partial x'}-\frac{(y_2-y_0)(y_2-y')}{(y_2-y_1)^2}\frac{\partial y_1}{\partial x'} \nonumber \\
&&-\frac{(y_1-y_0)(y'-y_1)}{(y_2-y_1)^2}\frac{\partial y_2}{\partial x'},\label{eq:partialderivyx}
\\
\frac{\partial y}{\partial y'}&=&\frac{y_2-y_0}{y_2-y_1}+\frac{y_2-y'}{y_2-y_1}\frac{\partial y_0}{\partial y'}-\frac{(y_2-y_0)(y_2-y')}{(y_2-y_1)^2}\frac{\partial y_1}{\partial y'} \nonumber \\
&&-\frac{(y_1-y_0)(y'-y_1)}{(y_2-y_1)^2}\frac{\partial y_2}{\partial y'}.\label{eq:partialderivyy}
\end{eqnarray}

\noindent Now after having derived the general formulas for mimesis, we turn to the numerical simulations. From formulas (\ref{eq:cylinders})-(\ref{eq:partialderivyy}), in order to construct our cloak, we only need to know {\bf \underline x}$_0$, {\bf \underline x}$_1$, {\bf \underline x}$_2$  and their respective derivatives.
The explicit  illustrations we have supplied to exemplify the work within this paper have boundaries whose horizontal plane sections are parts of ellipses (sunflower-like petal, cross, circle) or lines (parallelogram, hexagram, triangle).

\section{Mimetism for non-singular cloaks}

In Section \ref{chap:transformation} we presented the theoretical study of the mathematical model underlying our proposal for cloaks with any arbitrary star shaped boundaries cross sections, that perform mimetism as well as allowing invisibility. In the present section, we illustrate this by examples with piecewise linear or elliptic boundaries and provide their numerical validation.


 \subsection{Formulas for piecewise linear boundaries}\label{chap:linearboundaries}
 If a piece of the boundary of a star domain $D_i$ is part of a line of the form $y=a_ix+b_i,$ then clearly the line through the origin and a point $(x',y')$ intersects this piece of boundary at

\begin{eqnarray} (x_i,y_i)=(\frac{b_ix'}{y'-a_ix'}, \frac{b_iy'}{y'-a_ix'}) \label{eq:square}
\end{eqnarray}
 and hence
\begin{eqnarray}
\hspace{-2.5cm}
\frac{\partial x_i}{\partial x'}=\frac{b_iy'}{(y'-a_ix')^2}, ~ \frac{\partial x_i}{\partial y'}=-\frac{b_ix'}{(y'-a_ix')^2},~
\frac{\partial y_i}{\partial x'}=\frac{a_ib_iy'}{(y'-a_ix')^2}, ~
 \frac{\partial y_i}{\partial y'}=-\frac{a_ib_ix'}{(y'-a_ix')^2}.
\end{eqnarray}
Of course, in the case where this piece of boundary is a vertical segment with equation $x=c,$ then the above intersection is at  $(c, c\frac{y'}{x'}).$

 \subsection{Formulae for piecewise  elliptic boundaries}\label{chap:ellipses}
 We suppose here that a piece of at least one of our boundary curves $S_i$ is part of a nontrivial ellipse $\mathcal E_i$ with equation of the form $(x-a)^2/c_i^2+(y-b)^2/d_i^2$ where $(a,b)$ is our ventage point. Of course a line $(x(t),y(t))=(a,b) + t(x'-a,y'-b)$ passing through $(a,b)$ and a different point $(x',y')$  intersects $\mathcal E_i$ at two distinct points.
   For the construction, we need the point $(x_i,y_i)$ of that intersection which is nearer to $(x',y')$ in the sense that $x_i-a$ and $y_i-b$ have the same sign as $x'-a$ and $y'-b,$ respectively.

We have
\begin{eqnarray}
x_i&=&a+\frac{x'-a}{\sqrt{(x'-a)^2/c_i^2+(y'-b)^2/d_i^2}} \; , \nonumber\\
y_i&=&b+\frac{y'-b}{\sqrt{(x'-a)^2/c_i^2+(y'-b)^2/d_i^2}} \; .\label{eq:elliptic1}
\label{bleu}
\end{eqnarray}

This implies
\begin{eqnarray}
\frac{\partial x_i}{\partial x'} &=& \frac{1}{\sqrt{\frac{(x'-a)^2}{c_i^2}+\frac{(y'-b)^2}{d_i^2}}}-\frac{(x'-a)^2}{\left( \frac{( x'-a)^2}{c_i^2}+\frac{(y'-b)^2}{d_i^2} \right)^{\frac{3}{2}}c_i^2}, \\
\frac{\partial x_i}{\partial y'} &=& -\frac{(x'-a)(y'-b)}{\left( \frac{( x'-a)^2}{c_i^2}+\frac{(y'-b)^2}{d_i^2} \right)^{\frac{3}{2}}d_i^2}, \\
\frac{\partial y_i}{\partial x'} &=&
-\frac{(x'-a)(y'-b)}{\left( \frac{(x'-a)^2}{c_i^2}+\frac{(y'-b)^2}{d_i^2}\right)^{\frac{3}{2}}c_i^2}, \\
\frac{\partial y_i}{\partial x'} &=& \frac{1}{\sqrt{\frac{(x'-a)^2}{c_i^2}+\frac{(y'-b)^2}{d_i^2}}}-\frac{(y'-b)^2}{\left( \frac{( x'-a)^2}{c_i^2}+\frac{(y'-b)^2}{d_i^2} \right)^{\frac{3}{2}}d_i^2}.
\end{eqnarray}

\noindent We can then apply formulae (\ref{eq:cylinders})-(\ref{eq:partialderivyy}) to build any
generalized cloaks involving boundaries of elliptic types, where the center of the ellipse is the ventage point $(a,b)$.

\begin{figure}[h]
\hspace{3cm}
%
%
\scalebox{0.65}{\includegraphics{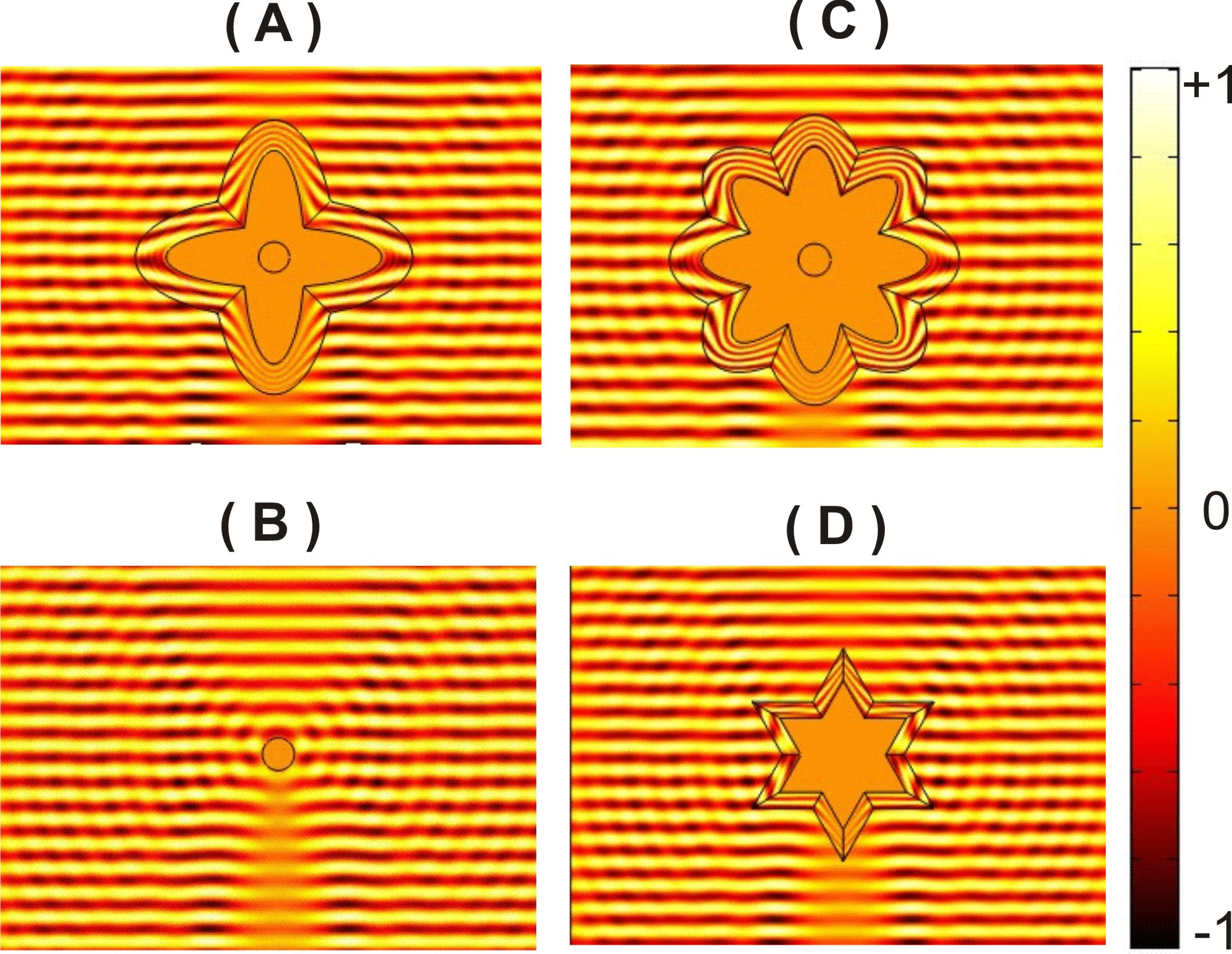}
}
\caption{\em\small~(Color online) A  cross-like (A), a sunflower-like (C) and a hexagram (D) all mimicking a circular cylinder (B) of small radius
$r_0=0.195$, that is $c_0=d_0=0.195$ in equation (\ref{bleu}).  The inner and outer boundaries of the petals are respectively parts of the ellipses $x^2/0.7^2+y^2/0.2^2=1$  and $x^2/0.9^2+y^2/0.4^2=1$ rotated by angle $0$ or $\frac{\pi}{2}$ in (A) or $0,$ $\frac{\pi}{4}$, $\frac{\pi}{2}$ or $3\pi/4$ in (C). The hexagram (D) is generated from an equilateral triangle with side $1.2.$ The energy corresponding to the plane matter
wave incident from the top is $E=\sqrt{\omega}=\sqrt{2\pi c/\lambda}=4.58$, where $\lambda=0.3$ is the wavelength of a transverse electromagnetic
wave in the optics setting with $c$ the celerity of light in vacuum, normalized here to $1$. To enhance the scattering,
a flat mirror is located under each quantum cloak and obstacle.} \label{fig:ellipse_as_circle1_petal4b}
\end{figure}

This has been used  in Figure  \ref{fig:ellipse_as_circle1_petal4b} (A), Figure  \ref{fig:ellipse_as_circle1_petal4b} (C),
Figures \ref{fig:squaring_the_circle} (E) and Figure  \ref{star} (B). We note that outside the cloaks in Figure  \ref{fig:ellipse_as_circle1_petal4b} (A) and  \ref{fig:ellipse_as_circle1_petal4b} (C), the scattered field is
exactly the same as that of the small disc of radius $r_0=0.195$. When the radius $r_0$ of the disc (that is,  when $c_0=d_0=r_0$ in equation (\ref{bleu})) tends to zero, the cloaks become singular and the plane matter wave goes unperturbed
(invisibility).

\begin{figure}[h]
\centerline{
\scalebox{0.80}{\includegraphics{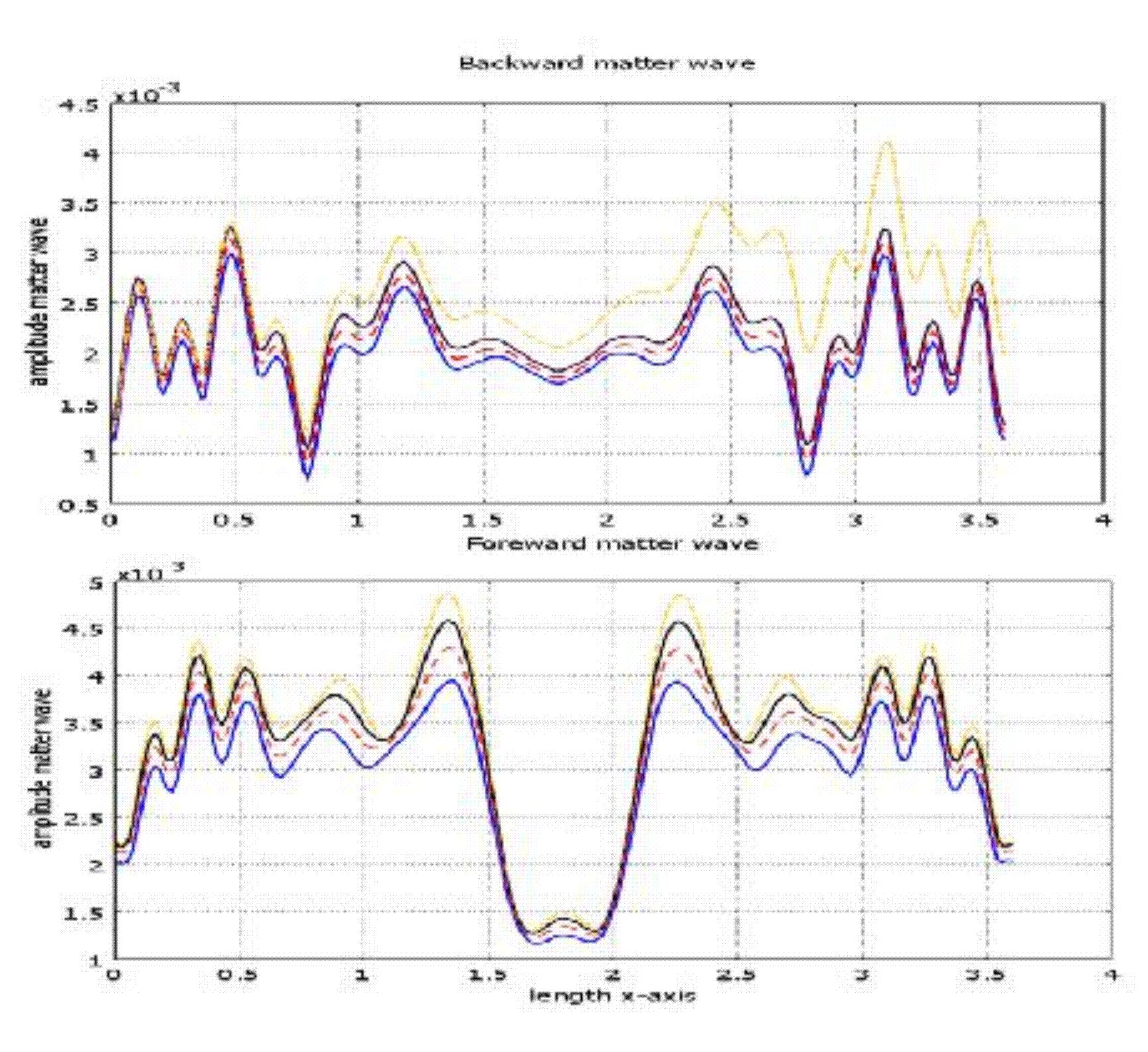}
}}
\caption{\em \small~(Color online) Upper panel: Profile of backward matter wave along the $x$-axis for $y=1$ for a plane wave
incident from the top as in Fig. \ref{fig:ellipse_as_circle1_petal4b} for a
cross-like (solid blue curve, see A), a sunflower-like (dashed red curve, see C), a hexagram (dotted yellow curve, see D) and a circular cylinder of small radius $r_0=0.195$ (solid black curve, see B);
Lower panel: idem for profile of foreward matter wave for $y=-1$; We note the large amplitude of the forward wave, due to the presence of a mirror below each nano scatterer at $y=-1.2$. The slight discrepancy between the curves is attributed to a numerical inaccuracy induced by the highly heterogeneous nature of the cloaks which is further enhanced by the irregularity of the boundary of the starshape cloak (dotted yellow curve, see D). The amplitude of the wave in both panels can be normalized to 1 (dividing throughout by $4.5 \, 10^{-3}$ and by $5 \, 10^{-3}$ in the upper and lower panels
respectively) for comparison with Fig. \ref{fig:ellipse_as_circle1_petal4b}.}
\label{fig:backfor}
\end{figure}




\begin{figure}[h]
\hspace{1cm}
\mbox{}
\scalebox{0.7}{\includegraphics{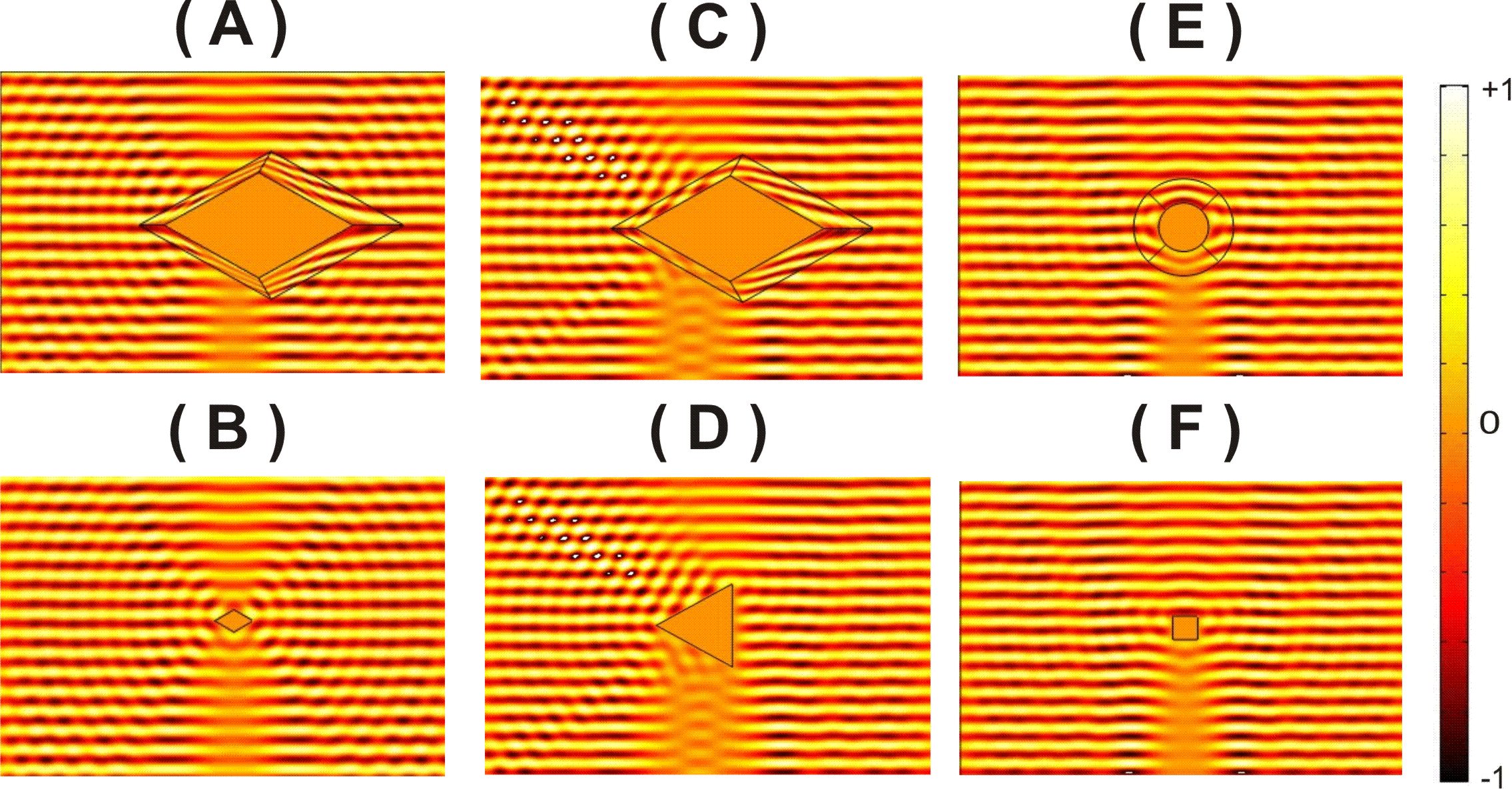}}
\vspace{-0.4cm}\caption{\em\small~(Color online) Left: A hollow parallelogram cylindrical region (A) scatters any incoming plane wave just like a much smaller solid cylinder (B) of the same nature. In (C) the same  hollow parallelogram cylinder is designed to have the same response to waves as the small equilateral triangle (D) with side $\frac{2\sqrt{3}}{5}$.
Right, squaring the circle: metamaterials allow to make a circular cylindrical hollow region (inner and outer radii $0.2$ and $0.4$ respectively) and a small solid square cylinder (of side $L_0=0.2$ and having the same center) equivalent, as regards their signatures and the way waves see them. In all these cases, the coated regions not only  gain the same electromagnetic signature as any desired other object, but also serve as cloaks with nonsingular material properties, in fact the presence of any types of defects hidden inside them has no effect in  the way they scatter waves. The energy corresponding to the plane matter wave incident from the top is $E=\sqrt{\omega}=\sqrt{2\pi c/\lambda}=4.58$, where $\lambda=0.3$ is the wavelength of a transverse electromagnetic
wave in the optics setting with $c$ the celerity of light in vacuum, normalized here to $1$.}
\label{fig:squaring_the_circle}
\end{figure}

\subsection{Squaring the circle}\label{chap:squarecircle}
In this section, we make a circle have the same (electromagnetic) signature as a virtual small square lying inside its enclosed region and sharing the same centre. In particular, its appearance to an observer will look like that of a square. As above, the transformation will map the region enclosed between the small square  and the outer circle into the circular annulus bounded by the inner (cloaking surface) and outer circles, where the sides of the square are mapped to the inner circle and the outer circle stays fixed point wise. To do so, we again use the diagonals of the square to part those regions into sectors. Indeed, the diagonals provide a natural triangulation by dividing the region inclosed inside the square into four sectors. The natural continuation of such a triangulation gives the needed one.
 In each sector,  we apply the same formulas  as above, where $(x_0,y_0)$ are obtained from the small square as in Section \ref{chap:linearboundaries} and for both $(x_1,y_1)$and $(x_2,y_2)$, we use the same formulas for ellipses in Section \ref{chap:ellipses}.
For the numerical simulation, we use a small square of side $L_0=0.2$
 and two circles radii $R_1=c_1=d_1=0.2$, $R_2=c_2=d_2=0.4$
all centred at $(a,b)=(0,0).$ So in both the uppermost and lowermost sectors (see Figure 3 (E) ) formulas (\ref{eq:square}) for the square become $ (x_0,y_0)=(L_0\frac{x'}{y'}, L_0), $  whereas in the leftmost and rightmost sectors, we have $ (x_0,y_0)=(L_0, L_0\frac{y'}{x'}).$ For all sectors, formulas (\ref{eq:elliptic1}) now read
$(x_i,y_i)=(R_ix'(x'^2+y'^2)^{-\frac{1}2},R_iy'(x'^2+y'^2)^{-\frac{1}2}),$ ~ $i=1,2.$

\noindent The inner boundary of the cloak corresponds to a potential wall (with transformed Neumann
boundary conditions, which also hold for infinite conducting or rigid obstacles depending upon the physical
context). We report these results in Figure \ref{fig:ellipse_as_circle1_petal4b} and Figure \ref{fig:squaring_the_circle}.
Some Neumann boundary conditions are set on the ground plane, the inner boundary of the carpet and the rigid obstacle.



\subsection{Star shaped cloaks}
In this section, we call star shaped a region bounded by a star polygon, such as a pentagram,
a hexagram, a decagram, ..., as in \cite{diattaguenneau-nz}.  Star shaped regions are particular cases of star regions.
The design of star shaped cloaks requires an adapted triangulation of the corresponding region. That is, a triangulation that takes into account
the singularities at the vertices of the boundary of the region. So that, each vertex belongs to the edge of some triangle. To the resulting triangles,
one applies the same maps as in Section \ref{chap:transformation}.
See e.g. Figure \ref{fig:ellipse_as_circle1_petal4b} (D).


\subsection{Finite Element Analysis of the cloak properties }
We now turn to specific numerical examples in order to illustrate the efficiency and feasibility of the cloaks we design.

\subsubsection{Comparison of backward and forward scattering of isomorphic cloaks}
Let us start with a comparison of both backward and forward scattering for the cloaks shown in Figure \ref{fig:ellipse_as_circle1_petal4b} (D).
We report in Figure \ref{fig:backfor} the amplitude of the matter wave above and below the scatter (cloak and/or obstacle) along the x-axis
respectively for $y=1$ and $y=-1.2$. We note the slight discrepancy between the curves, which is a genuine numerical artefact: we have
checked that the finer the mesh of the computational domain, the smaller the discrepancy (which is a good test for the convergence of the
numerical package COMSOL Multiphysics). The mesh needs actually be further refined within the heterogeneous anisotropic cloak and the
perfectly matched layers compared with the remaining part of the computational domain which is filled with isotropic homogeneous material. We note that the
yellow curve (corresponding to the pentagram, see Figure \ref{fig:ellipse_as_circle1_petal4b} D) is most shifted with respect
to the black curve (corresponding to the small obstacle on its own i.e. the benchmark, see  Figure \ref{fig:ellipse_as_circle1_petal4b} B).
This can be attributed to the irregular boundary of the cloak as analysed in the case of singular star-shaped cloaks in \cite{diattaguenneau-nz}:
we considered around 70000 elements for the mesh in all four computational cases reported in Figure \ref{fig:backfor} in order to exemplify the numerical
inaccuracies. For computations with 100000 elements, the yellow curve is shifted downwards and is nearly superimposed with the black curve.
Moreover, the strong asymmetry of the yellow curve in the upper panel of  Figure \ref{fig:backfor} vanishes for 100000 elements. We attribute
this numerical effect to the artificial anisotropy induced by the triangular finite element mesh of the pentagram.

\subsubsection{Analysis of the material parameters within a cloak}
 Another interesting point in this paper, is that the metamaterial in our model is non-singular, therefore enabling the implementation  of  potentially broadband cloaks
over a wide range of wavelengths. Moreover, the cloaked objects display a mimesis phenomenon in that, they are designed to acquire any desired quantum or
electromagnetic signature.

\begin{figure}[h]
\hspace{2cm}\scalebox{0.35}{
\includegraphics[width=40cm,angle=0]{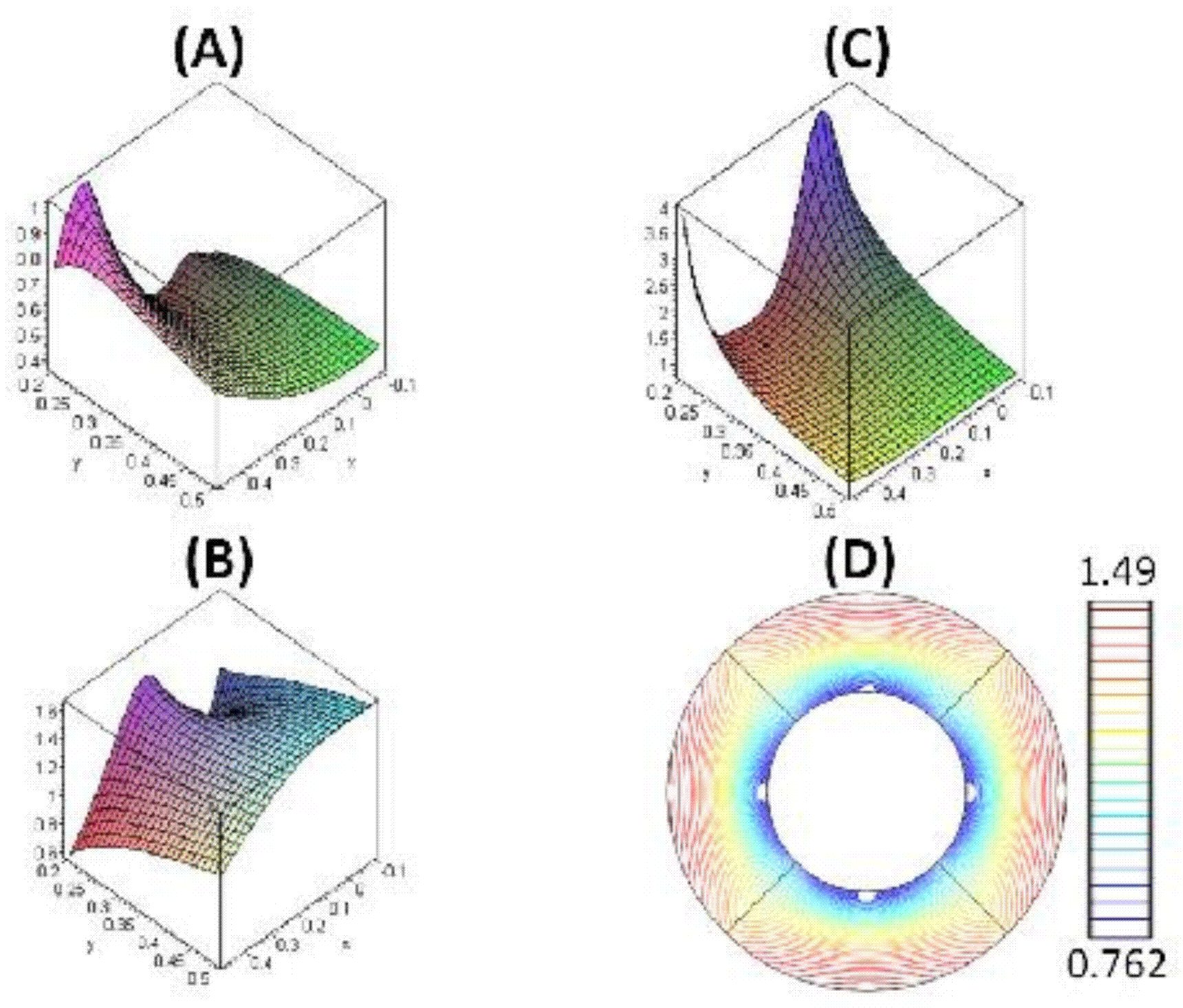}}
\caption{\em\small~(A)-(C) are illustration of the graphs $\{(x,y,\Lambda_i^{-1})\},$~ $i=1,2,3$ of the  three eigenvalues $\Lambda_i^{-1}$ of the material tensor ${\bf T}^{-1},$ in the uppermost sector of Fig. \ref{fig:squaring_the_circle} (E):
(A) $\Lambda_1^{-1}$; (B) $\Lambda_3^{-1}$; (C) $\Lambda_2^{-1}$.
We note that, each of all those three surfaces are strictly above the plane $z=0$ even inside the cloak itself.
Because all other sectors of the cloak are obtained by a rotation of the uppermost one, it suffices to study the eigenvalues of ${\bf T}^{-1}$  in just one sector. Note that, the above surfaces were drawn using the MAPLE software. In (D) we represent the finite element computation of  $\lambda_3$ (see B) for all four sectors of  Fig. \ref{fig:squaring_the_circle} (E) in the COMSOL MULTIPHYSICS package. We note the four-fold symmetry of the isovalues.}
\label{coeffmaterial}
\end{figure}

The result of the numerical exploration of the three eigenvalues $\Lambda_i^{-1}(x,y)$ of the material tensor ${\bf T}^{-1},$ of Figure \ref{fig:squaring_the_circle} (E), is reported in Figure \ref{coeffmaterial}. First, using Maple software, we have purposedly represented those eigenvalues in a wider domain, including the cloaked region, despite the boundary conditions on the inner boundary of cloak. One  clearly sees that all of the three eigenvalues take finite values away from $0$ even inside the cloak itself.
The material tensor ${\bf T}^{-1}$ is hence nonsingular. Due to the fourfold symmetric geometry of the cloak (all sectors are obtained by a rotation of any fixed one), we only need to look at those eigenvalues in one sector. We note that $4>\Lambda_2^{-1}>\Lambda_1^{-1}>0.2$ in accordance with the fact that the cloak should display a strong anisotropy in the azimuthal direction in order to detour the wave. For a singular cloak $\Lambda_1^{-1}$ tends to zero on the inner
boundary of the cloak, while $\Lambda_2^{-1}$ tends to infinity.

We also represent the finite element simulation of  $\Lambda_3^{-1}$, which exemplifies the fourfold symmetry of the
isovalues within the circular cloak, a fact reminiscent of the fourfold symmetry of the square which the cloak mimics.

\section{Generalized mirage effect and almost trapped states}
It is known that a point source located inside the coating of a singular cloak
(i.e. a cloak such that $x_0=y_0=0$ in (\ref{eq:cylinders}) leads to a mirage effect
whereby it seems to radiate from a shifted location in accordance with this
geometric transformation \cite{zolla_oplett07}.

\subsection{Shifted quantum dot inside the transformed space}
\begin{figure}[h]
\hspace{2cm}\scalebox{0.35}{
\includegraphics[width=40cm,angle=0]{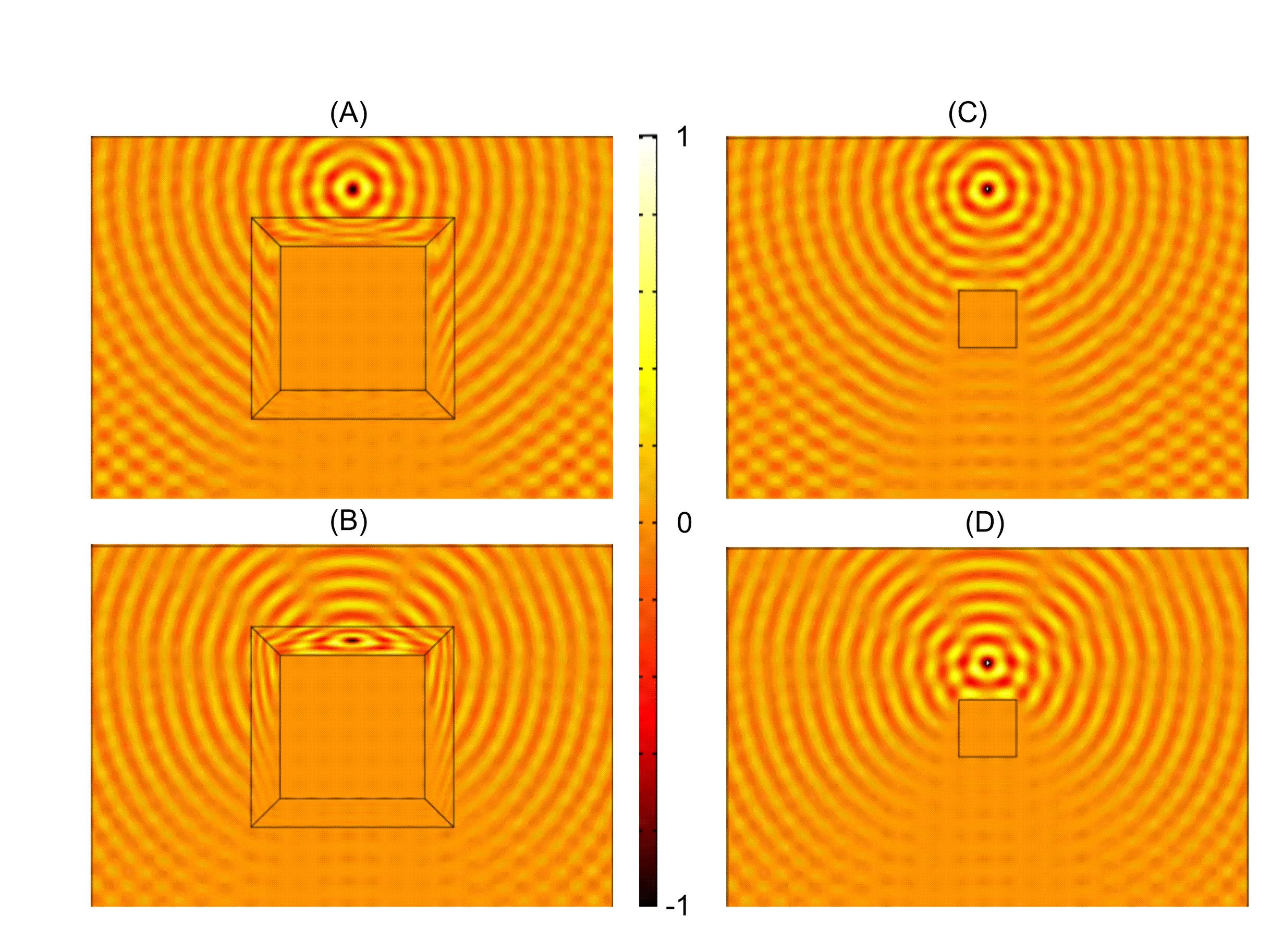}}
\mbox{}\vspace{-1cm}\caption{\em\small~A non-singular cloak with
two square boundaries of sidelengths $1$ and $1.4$ in presence of
a quantum dot with energy $E= 4.58$ (resp. an electric current line source of wavelength $\lambda=0.3$ in the
context of tranverse electric waves).
(A-C) When the source is located a distance $0.2$ away from the cloak,
it seems to emit as if it would be a distance $0.4$ away from a square obstacle
of sidelength $0.4$. (b-d) When the source is located a distance $0.1$ away from the inner boundary
of the cloak (i.e. in the middle of the coating), it seems to emit as if it would be a distance $0.25$ away from a square obstacle
of sidelength $0.4$, in accordance with (\ref{eq:cylinders}) where $x_0=0.2$, $x_1=0.5$ and $x_2=0.7$.
}\label{mirage}
\end{figure}

This prompts the question as to whether a similar effect can be
observed in non-singular cloaks i.e. when $x_0$ or $y_0$ are nonconstant. As it turns out, the physics
is now much richer: we can see in Figure \ref{mirage} that when the source lies
inside the coating, it only seems to radiate from a shifted location
in accordance with (\ref{eq:cylinders}), but it is moreover in presence
of a small object of sidelength $2x_0$. This can be seen as a generalized
mirage effect which opens many new possibilities in optical illusions. Indeed, Nicolet
et al. have proposed to extend the concept of mirage effect for point sources located
within the heterogeneous anisotropic coating of invisibility cloaks to finite size
bodies which scatter waves like bodies shaped by the geometric transform
\cite{nicolet2010a}. This is in essence an alternative path to our proposal for
mimetism. However, this mirage effect can be further generalized to non-singular
for which a finite size body located inside the coating will now create the optical
illusion of being another body in presence of some obstacle, which bears some
resemblance with Fata Morgana, a mirage which comprises several inverted (upside down)
and erect (right side up) images that are stacked on top of one another. Such a mirage occurs
because rays of light are bent when they pass through air layers of different temperature.
This creates the optical illusion of levitating castles over seas or lakes, as reported by
a number of Italian sailors, hence the name related to Morgana, a fairy central to the
Arthurian legend able to make huge objects fly over her lake.

\subsection{Field confinement on resonances: Anamorphism fall down}
\begin{figure}[h]
\hspace{2cm}\scalebox{0.35}{
\includegraphics[width=40cm,angle=0]{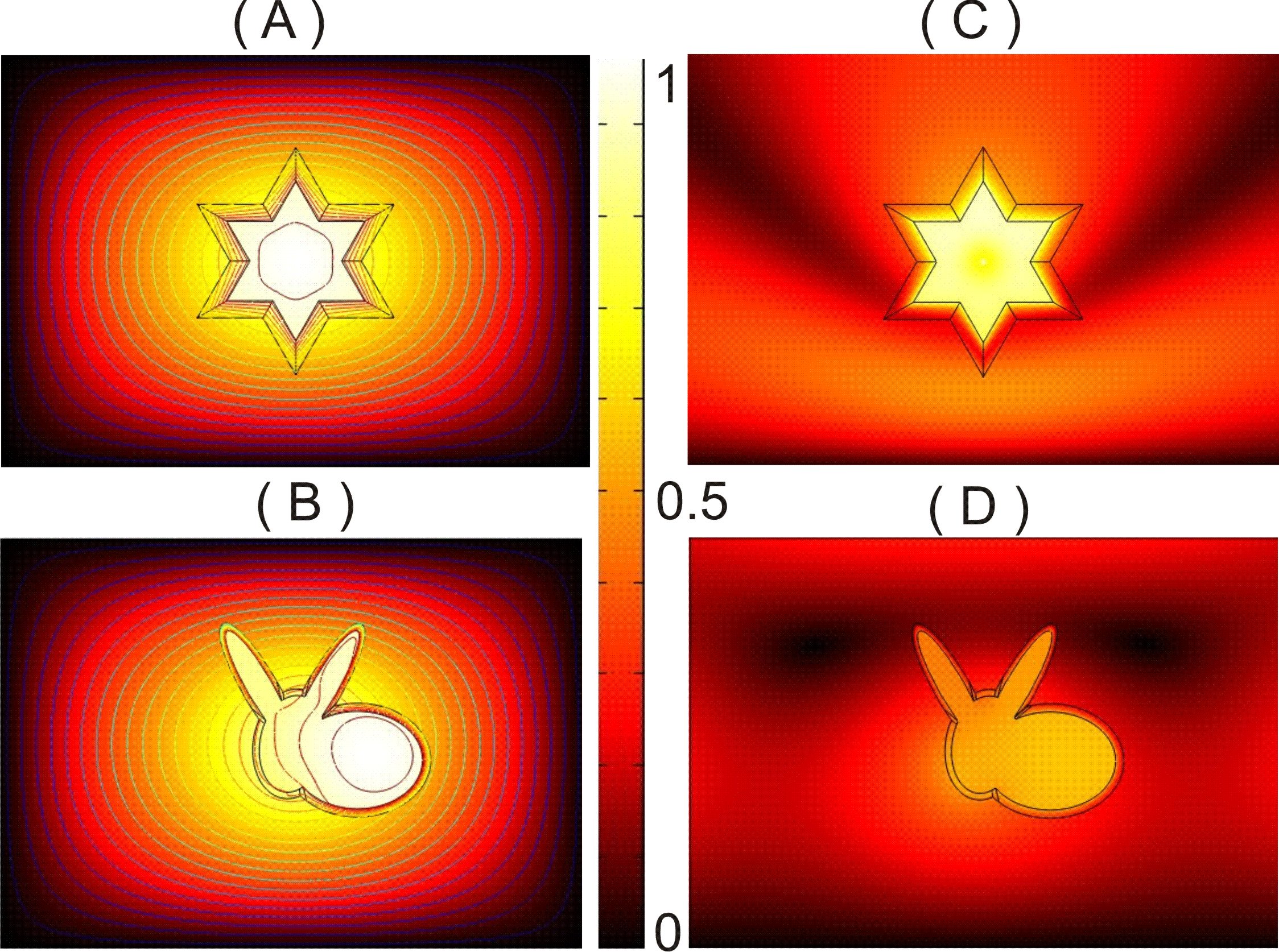}}
\mbox{}\vspace{0cm}\caption{\em\small~Left panel: Modulus of the fundamental eigenstates associated with quantified normalized energy $E=\sqrt{\omega}=\sqrt{2\pi c/\lambda}=1.32$
(resp. a transverse electric plane wave of wavelength $\lambda=3.6$)
for a non-singular cloak shaped as a star (A) and a rabbit (B) both
of which mimic a small disc (of normalized radius $0.195$, e.g. $195$ nanometers) ; Right panel: Matter wave incident from the top on the quantum
cloaks with a spatially varying potential $V'$ with compact support (i.e. vanishing outside the cloak) and energy $E$;
The large amplitude of the field within the cloak is noted.}
\label{trapped}
\end{figure}

Another intriguing feature of singular cloaks is their potential for light confinement
associated with almost trapped states which are eigenfields exponentially decreasing
outside the invisibility region. Such modes were described in the context of quantum cloaks
by Greenleaf, Kurylev, Lassas and Uhlmann in \cite{greenleaf-trapped}. These researchers
discovered that such modes are associated with energies for which the Dirichlet to Neunmann
map is not defined i.e. on a discrete set of values. Here, we revisit their paper in light
of non-singular quantum cloaks, that is when we consider the blowup of a small region instead
of a point. Our findings reported in Figure \ref{trapped} for a star shaped and a rabbit-like
non-singular cloaks mimicking a small disc of radius $0.195$ bridge the quantum mechanical
spectral problems (panels (a) and (b)) to the scattering problems (panels (c) and (d))
in the following way: we first look for eigenvalues (i.e. quantified energies $E$) and
associated eigenfunctions $\Psi$ of the equation Eq. (\ref{transfpressure}) in the class
of square integrable functions on the whole space $\R^2$ (note that here, as the metric is
non singular, there is no need to consider a weighted Sobolev space). Note however that
the set continuity conditions on the inner boundary of the cloak, instead of Neumann ones.
This provides us with
a discrete set of complex eigenfrequencies, with a very small imaginary part (also known as
leaky modes in the optical waveguide literature). We neglect this imaginary part
and launch a plane wave on the non-singular cloak (whereby the invisibility region is also
included within the computational domain as we once again set continuity conditions on the inner
boundary of the cloak) at the very frequency given by the spectral problem, see panels (c) and (d).
We clearly see that the inside of the cloak hosts a quasi-localized eigenstate whose energy is
mostly confined inside a star (panel C) and a rabbit (panel D), both of which actually
scatter like a small disc.

\section{Generalized super-scattering for negatively refracting non-singular cloaks}
In this section, we take some freedom with the one-to-one feature of the previous transforms
and allow for space folding. This means that $|x_0|\ge |x_2|\ge |x_1|$ and $|y_0|\ge |y_2|\ge |y_1|$ in (\ref{eq:cylinders})
while the $x_i$ (resp. the $y_i$)  i=1,2,3 all have the same sign,
thus making $\alpha$ and $\beta$ strictly negative real-valued functions. It has been known for a while
that space folding allows for the design of perfect lenses, corners and checkerboards \cite{sar1,sar2,mknjp2010,ulf2006a}. But it is only
recently that researchers foresaw the very high-potential of space folding as applied to the
design of super-scatterers \cite{tunnel,chan2009a,chan2009b,chan2009c,weependry2010}. We generalize these concepts to mimetism
via space folding.

\begin{figure}[h]
\scalebox{0.4}{
\includegraphics[width=40cm,angle=0]{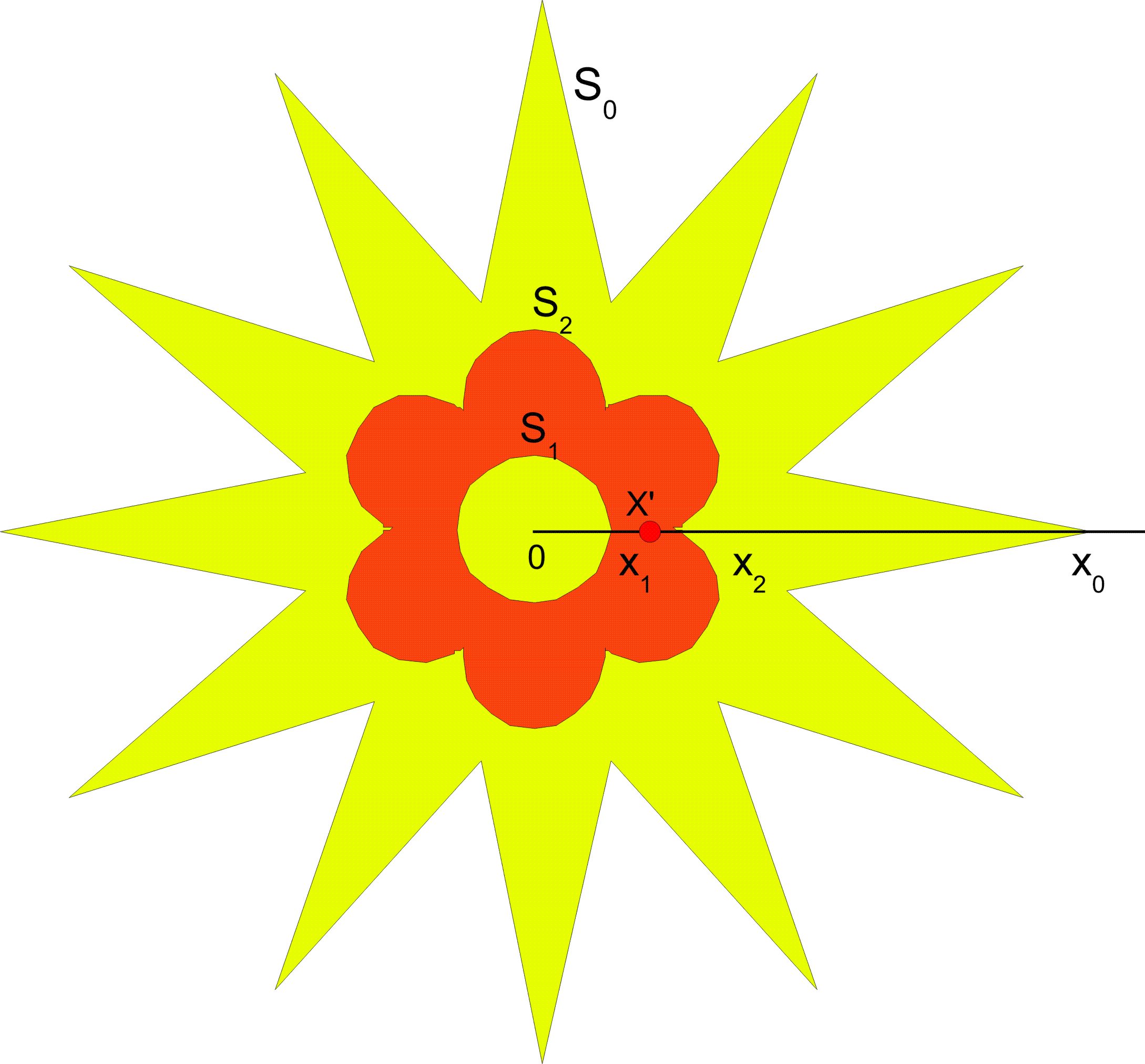}}
\mbox{}\vspace{-0.4cm}\caption{\em\small~Construction of a generalized cloak with optical space folding for superscattering effect.
The transformation  with inverse (\ref{eq:cylinders}) magnifies the
region bounded by the two surfaces $S_0$ and
$S_2$ into the region bounded by $S_1$
and $S_2$. We note that the transform is no longer an isomorphism.
The curvilinear  metric inside the carpet (here, an orange flower) is
described by the transformation matrix ${\bf T},$ see (\ref{eq:tensorgeneral})-(\ref{eq:partialderivyy}).
Any quantum object located within the region bounded by the surface $S_1$ scatters matter waves like
a larger object bounded by $S_0$ (here, a yellow star).}\label{fig:constructris}
\end{figure}

\noindent The mapping leading to the super-scatterer is shown in figure \ref{fig:constructris}.
We would like to emphasize that here we get not only a magnification of the
scattering cross section of an object, in a way similar to what optical space folding
does for a cylindrical perfect lens, but we can also importantly change
the shape of the object.

\begin{figure}[h]
\hspace{2cm}\scalebox{0.35}{
\includegraphics[width=40cm,angle=0]{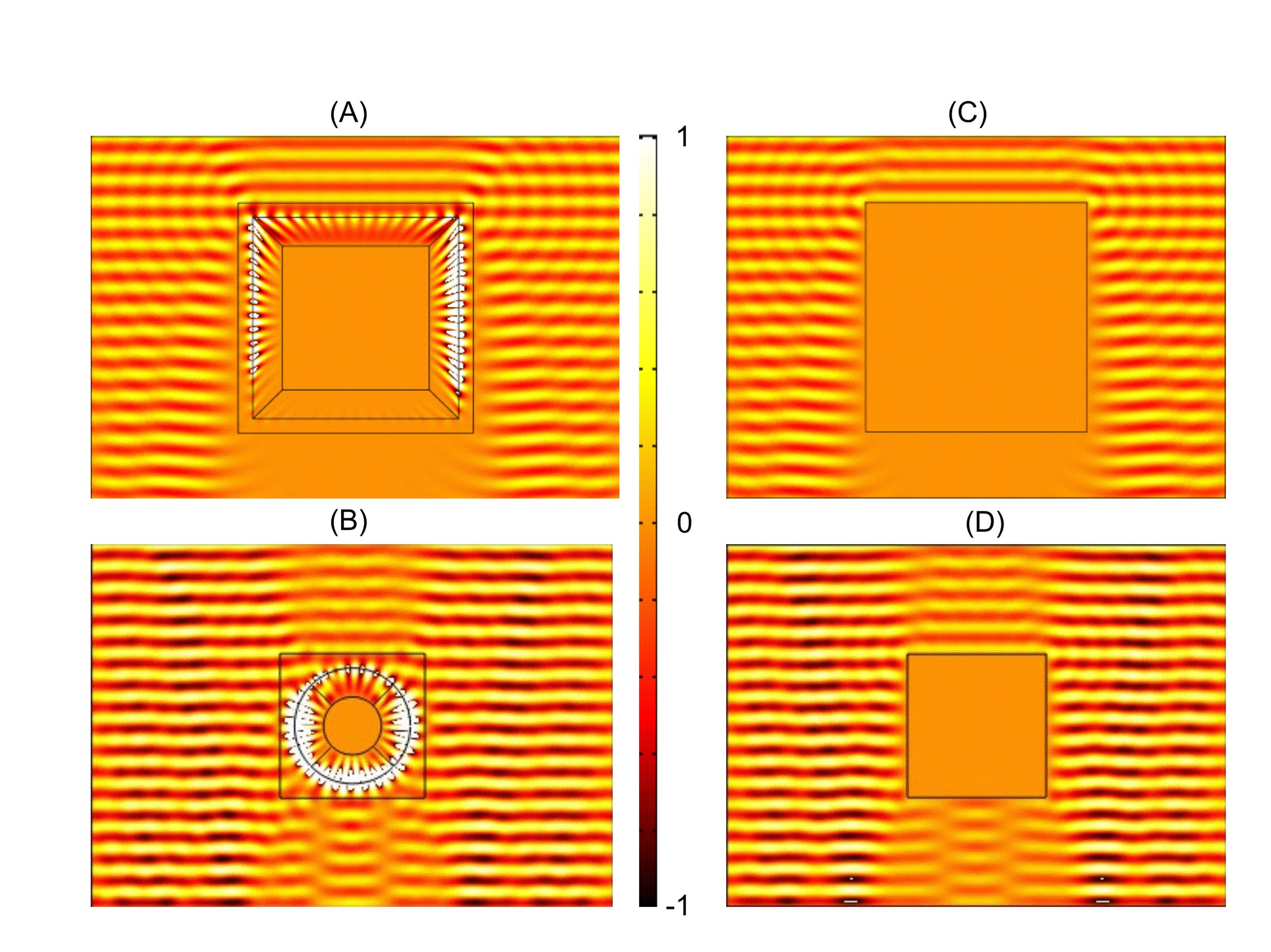}}
\mbox{}\vspace{-1cm}\caption{\em\small~(A-C)Any obstacle surrounded by a square anti-cloak with
square boundaries of sidelengths $1$ and $1.4$,
scatters a plane matter wave of energy $E=4.58$ (resp. a transverse electric plane wave of wavelength $\lambda=0.3$ in the
context of optics) which is coming from the top like a larger square obstacle
of sidelength $1.6$.
(b-d) Any obstacle surrounded by a circular anti-cloak with circular boundaries of radii $0.2$ and $0.4$
scatters a plane matter wave of energy $E=4.58$ (resp. a transverse electric plane wave of wavelength $\lambda=0.3$)
which is coming from the top like a larger square obstacle of sidelength $1$.
The large field amplitude on the upper boundary of the anti-cloak in (A) and (B) is noted.
It can be attributed to some kind of surface matter wave (a surface plasmon polariton in the context of optics).}
\label{star}
\end{figure}

We illustrate our proposal with a numerical simulation for a square obstacle surrounded by
an anti-cloak in Figure \ref{star}(a), which mimics a larger square obstacle,
see Figure \ref{star}(b). We note that the large field amplitude on the anti-cloak's
upper boundary can be attributed to a surface field arrising from the physical
parameters with opposite signs on the cloak outer boundary
(an anisotropic mass density in the context of quantum mechanics and an anisotropic
permittivity in the context of optics).
It is illuminating here to draw some correspondence with electromagnetic waves, as
the anisotropic mass density (resp. permittivity) indeed takes opposite values when we cross the outer
boundary of the cloak, and this ensures the existence of a surface matter wave (resp. a surface
plasmon polariton) clearly responsible for the large field amplitude (in the transverse electric wave
polarization i.e. for a magnetic field parallel orthogonal to the computational plane.
We further show an example of a small circular obstacle mimicking a large square
obstacle in Figure \ref{star}(a) and \ref{star}(b). Once again, the large field amplitude
on the outer cloak boundary comes from the complementary media inside and
outside the cloak. We believe such types of mimetism might have tremendous
applications in quantum dot probing, bringing the nano-world a step closer to metamaterials.



\section{Conclusion}
In this paper, we have proposed some models of generalized cloaks that create
some illusion. We focussed here on the Schr\" odinger equation which is valid
in a number of physical situations, such as matter waves in quantum optics.
However, the results within this paper can easily be extended to the
Helmholtz equation which governs the propagation of acoustic and
electromagnetic waves at any frequency. One simply needs to insert
the transformation matrix within the shear modulus and density
of an elastic bulk (in the case of anti-plane shear waves), the density and compressional
modulus of a fluid (in the case of pressure waves), or the permittivity and permeability
of a medium (in the case of electromagnetic waves) \cite{milton,cummernjp,sanchez,pendryprl,chen07,farhat_prl}.
In this latter context, this means that in transverse electric polarization (whereby the magnetic field is parallel
to the fiber axis), infinite conducting obstacles dressed with these cloaks
display an electromagnetic response of other infinite conducting obstacles. In these cloaks,
an electric wire could in fact be hiding a larger object near it. Actually, any object
could mimic the signature of any other one. For instance, we design a cylindrical cloak so that a circular obstacle behaves like
a square obstacle, thereby bringing about one of the oldest
enigma of ancient time : squaring the circle! The ordinary singular cloaks then come as a particular
case, whereby objects appear as an infinitely small infinite conducting object
(of vanishing scattering cross-section) and hence
become invisible. On the contrary, such generalized cloaks are described by non-singular permittivity and
permeability, even at the cloak's inner surface. We
have proposed and discussed some interesting applications in the context of quantum mechanics such as
probing nano-objects.

\noindent Obviously, one realizes that, when the inner virtual region $D_0$ responsible for mimetism
tends to zero, one recovers
the case in \cite{pendrycloak} where the material properties are no
longer bounded, as one of the eigenvalues of the mass density matrix tends to zero whilst another one
recedes to infinity, as we approach the inner boundary of the coated
region, see also \cite{kohn}.

In this paper, we have played with optical illusions, trying to be as imaginative as possible in order to exhaust the possible
geometric transforms we had at hand in Euclidean spaces (Non-Eulidean cloaking  is a scope for more creative thinking \cite{nons1}).
It should be pointed out that while the emphasis of this paper was on quantum waves, corresponding non-singular cloaks in electromagnetism
that have an inner boundary which is perfectly electric conducting (PEC) and scatter like a reshaped PEC object, were investigated in
\cite{nonsing1,nonsing2,nonsing3,nonsing3,nonsing4,nonsing5,nonsing6}. However these works focussed mostly on the reduction of the
scattering cross section of a diffracting object, while the present paper explores the mimetism effect whereby an object scatters like another
object of any other scattering cross section (and in particular reduced or enhanced ones).

Metamaterials  \cite{zheludev} is a vast area with a variety of composites structured on the subwavelength scale in order to sculpt the electromagnetic wave trajectories, as experimentally demonstrated
at microwave frequencies by a handful of research groups worldwide \cite{cloakex,kante,tretyakov}. Resonant elements within metamaterials are in essence man-made atoms allowing
to mimic virtually any electromagnetic response we wish, and this is turn allows us to push
the frontiers of photonics towards previously unforeseen areas.

\section*{Acknowledgements}
AD and SG acknowledge funding from EPSRC grant EPF/027125/1. We authors
also wish to thank the anonymous referees for constructive critical comments.



\section*{References}
\begin{thebibliography}{99}
\bibitem{pendrycloak} Pendry J B, Shurig D and Smith D R 2006
"Controlling electromagnetic fields,"
Science {\bf 312} 1780-1782.

\bibitem{ulf2006a}
Leonhardt U 2006
"Optical conformal mapping,"
Science {\bf 312}, 1777-1780

\bibitem{ulf2006b}
Leonhardt U
and Philbin T G 2006
"General Relativity in Electrical Engineering,"
New J. Phys. {\bf 8}, 247

\bibitem{veselago} Veselago V G 1967 Usp. Fiz. Nauk {\bf 92} 517-526

\bibitem{veselago68} Veselago V G 1968 Sov. Phys.�Usp. {\bf 10} 509-514

\bibitem{pendry_prl00} Pendry J B 2000
"Negative refraction makes a perfect lens,"
Phys. Rev. Lett. {\bf 86}, 3966-3969.

\bibitem{smith00}
Smith D R, Padilla W J, Vier V C, Nemat-Nasser S C and Schultz
S  2000 Phys. Rev. Lett. {\bf 84}, 4184-4187

\bibitem{sar_rpp05}
Ramakrishna S A 2005 Rep. Prog. Phys. {\bf 68}  449-521

\bibitem{cheianov}
Cheianov V V, Fal'ko V, Altshuler B L 2007 Science {\bf 315}, 1252-1255

\bibitem{moreno} Moreno E, Fern\'andez-Dom\'inguez A I, Cirac J I, Garc\'ia-Vidal F J and Mart\'in-Moreno L 2005 Phys. Rev. Lett. {\bf 95} 1704061-1704064

\bibitem{milton} Milton G W, Briane M and Willis J R 2006 New J. Phys. {\bf 8} 248

\bibitem{cummernjp} Cummer S A and Schurig D 2007 New J. Phys. {\bf 9} 45

\bibitem{sanchez} Torrent D and Sanchez-Dehesa J 2008 New J. Phys. {\bf 10} 023004

\bibitem{pendryprl} Cummer S A, Popa B I, Schurig D, Smith D R, Pendry J,
Rahm M and Starr A 2008 Phys. Rev. Lett. {\bf 100} 024301

\bibitem{chen07} Chen H and Chan C T
2007 Appl. Phys. Lett. {\bf 91} 183518

\bibitem{farhat_prl} Farhat M, Enoch S, Guenneau S and Movchan A B 2008
"Broadband cylindrical acoustic cloak for linear surface waves in a fluid,"
Phys. Rev. Lett. {\bf  101} 134501

\bibitem{apl2009}
Brun M, Guenneau S and Movchan A B 2009
"Achieving control of in-plane elastic waves,"
Appl. Phys. Lett. {\bf 94} 061903

\bibitem{prb2009}
Farhat M, Guenneau S, Enoch S and Movchan AB 2009
"Cloaking bending waves propagating in thin plates,"
Phys. Rev. B {\bf 79} 033102

\bibitem{matter_prl2008}
 Zhang S, Genov D A, Sun C and Zhang X 2008
"Cloaking of matter waves,"
Phys. Rev. Lett. {\bf 100} 123002

\bibitem{greenleaf-trapped} Greenleaf A, Kurylev Y, Lassas M, Uhlmann G 2008
New J. Phys. {\bf 10} 115024

\bibitem{kohn}
Kohn R V, Shen H, Vogelius M S and Weinstein M I 2008
"Cloaking via change of variables in electric impedance tomography,"
Inverse Problems {\bf 24} 015016

\bibitem{greenleaf1} Greenleaf A, Kurylev Y, Lassas M, Uhlmann G 2007
"Full-wave invisibility of active devices at all frequencies,"
Comm. Math. Phys. {\bf 275}(3) 749-789

\bibitem{zolla_oplett07} Zolla F, Guenneau S, Nicolet A and Pendry J B 2007
"Electromagnetic analysis of cylindrical invisibility cloaks and mirage effect,"
Opt. Lett. {\bf 32} 1069-1071

\bibitem{nicolet2010a}
Nicolet A, Zolla F, and Geuzaine C,
"Generalized Cloaking and Optical Polyjuice,"
ArXiv:0909.0848v1.

\bibitem{Ross_cloaking} Nicorovici N A, McPhedran R C and Milton G W 1994
"Optical and dielectric properties of partially resonant composites,"
Phys. Rev. B {\bf 49} 8479-8482

\bibitem{torres} Torres M, Adrados J P, Montero de Espinosa F R, Garcia-Pablos D,
and Fayos J 2000 Phys. Rev. E {\bf 63} 011204

\bibitem{greenleaf}
Greenleaf A, Lassas M and Uhlmann G 2003
"On nonuniqueness for Calderon's inverse problem,"
Math. Res. Lett. {\bf 10} 685-693

\bibitem{nons1}
Leonhardt U and Tyc T 2008,
"Broadband invisibility by an euclidean cloaking,"
Science {\bf 323}(5910), 110-112
\bibitem{nons2}
Zhang P, Jin Y, and He S 2008
"Obtaining a nonsingular two-dimensional cloak of complex shape from a perfect three-dimensional cloak,"
Appl. Phys. Lett. {\bf 93}, 243502-243504

\bibitem{nons2}
Collins P and McGuirk J 2009
"A novel methodology for deriving improved material parameter sets for simplified cylindrical cloaks,"
J. Opt. A: Pure Appl. Opt. {\bf 11}, 015104-015111

\bibitem{smithcarpet} Liu R, Ji C, Mock J J, Chin J Y, Cui T J and Smith D R 2008
"Broadband Ground-Plane Cloak,"
Science {\bf 323} 366-369

\bibitem{pendry-carpet} Li J and Pendry J B 2008
"Hiding under the Carpet: A New Strategy for Cloaking,"
Phys. Rev. Lett. {\bf 101} 203901-4

\bibitem{pendrynjp}Pendry J B and Li J 2008 New J. Phys. {\bf 10} 115032

\bibitem{norris} Norris A N 2008 Proc. Roy. Soc. Lond. A {\bf 464} 2411-2434

\bibitem{njpnico}Nicorovici N A P, McPhedran R C, Enoch S and Tayeb G 2008
 "Finite wavelength cloaking by plasmonic resonance,"
New J. Phys. {\bf 10} 115020

\bibitem{nicolae}
Milton G and Nicorovici NA 2006
"On the cloaking effects associated with anomalous localized resonance,"
Proc. Roy. Soc. Lond. A {\bf 462}, 3027

\bibitem{alu2005}
Alu A and Engheta N 2005
"Achieving Transparency with Plasmonic and Metamaterial Coatings,"
Phys. Rev. E {\bf 95}, 016623

\bibitem{greenleaf-handlbdy} Greenleaf A, Kurylev Y, Lassas M, Uhlmann G 2008
"Electromagnetic wormholes via handlebody constructions,"
Comm. Math. Phys. {\bf 281} (2) 369-385

\bibitem{gabrielli-carpet} Gabrielli L H, Cardenas J, Poitras C B and Lipson M 2009
"Silicon nanostructure cloak operating at optical frequencies,"
 Nature Photonics {\bf 3} 461-463

\bibitem{diattaguenneau} Diatta A, Guenneau S, Dupont G and Enoch S 2010
"Broadband cloaking and mirages with flying carpets,"
Opt. Express {\bf 18} 11537-11551

\bibitem{diattaguenneau-nz} Diatta A, Nicolet A, Guenneau S and Zolla F   2009
"Tessellated and stellated invisibility,"
Opt. Express {\bf 17} 13389-13394

\bibitem{sar1}
Pendry JB and Ramakrishna SA 2003
"Focussing light with negative refractive index,"
J. Phys.: Condens. Matter {\bf 15}, 6345

\bibitem{sar2}
Guenneau S, Vutha AC and Ramakrishna SA 2005
 "Negative refractive in checkerboards related by mirror-antisymmetry and 3-D corner reflectors,"
New J. Phys. {\bf 7}, 164

\bibitem{mknjp2010}
Milton GW, Nicorovici NAP, McPhedran RC, Cherednichenko K and Jacob Z 2008
"Solutions in folded geometries, and associated cloaking due to anomalous resonance,"
New J. Phys. {\bf 10}, 115021

\bibitem{tunnel}
Zhang JJ, Luo Y, Chen SH, Huangfu J, Wu BI, Ran L and Jong JA 2009
"Guiding waves through an invisible tunnel,"
Opt. Express {\bf 17}, 6203

\bibitem{chan2009a}
Lai Y, Chen H, Zhang ZQ and Chan CT 2009
"Complementary Media Invisibility Cloak that Cloaks Objects at a Distance Outside the Cloaking Shell,"
Phys. Rev. Lett. {\bf 102}, 093901

\bibitem{chan2009b}
Lai Y, Ng J, Chen HY, Han DZ, Xiao JJ, Zhang ZQ and Chan CT 2009
"Illusion Optics: The Optical Transformation of an Object into Another Object,"
Phys. Rev. Lett. {\bf 102}, 253902

\bibitem{chan2009c}
Ng J, Chen HY and Chan CT 2009
"Metamaterial frequency-selective superabsorber,"
Opt. Lett. {\bf 34}, 644

\bibitem{weependry2010}
Wee WH, Pendry JB 2010
"Super phase array,"
New J Phys. {\bf 12} 033047

\bibitem{zheludev}
Zheludev NI 2010 Science { \bf 328}, 582

\bibitem{cloakex} Schurig D, Mock J J, Justice B J, Cummer S A, Pendry J B, Starr A F, Smith D R 2006
"Metamaterial electromagnetic cloak at microwave frequencies,"
Science {\bf 314} 977-980

\bibitem{kante}
Kante B, Germain D, de Lustrac A 2009
"Experimental demonstration of a nonmagnetic metamaterial cloak at microwave frequencies,"
Phys. Rev. B {\bf  80} 201104

\bibitem{tretyakov}
Tretyakov S, Alitalo P, Luukkonen O, Simovski C 2009
"Broadband electromagnetic cloaking of long cylindrical objects,"
Phys. Rev. Lett. {\bf 103} 103905

\bibitem{nonsing1}
Cummer SA, Rupoeng L and Cui TJ 2009
"A rigorous and nonsingular two dimensional cloaking coordinate transformation,"
Jour. Appl. Phys. {\bf 105}, 056102

\bibitem{nonsing2}
Hu J, Zhou X and Hu G 2009
"Nonsingular two dimensional cloak of arbitrary shape,"
Appl. Phys. Lett {\bf 95}, 011107

\bibitem{nonsing3}
Jiang WX, Cui TJ, Yang XM, Cheng Q, Liu R and Smith DR 2008
"Invisibility cloak without singularity,"
Appl. Phys. Lett. {\bf 93}, 194102

\bibitem{nonsing4}
Chen H, Zhang X, Luo X, Ma H and Chan CT 2008
"Reshaping the perfect electrical conductor cylinder arbitrarily,"
New J. Phys. {\bf 10}, 113016

\bibitem{nonsing5}
Li C, Yao K and Li F 2008
"Two-dimensional electromagnetic cloaks with non-conformal inner and outer boundaries,"
Opt. Express {\bf 16}(23), 19366

\bibitem{nonsing6}
Jiang WX, Ma HF, Cheng Q and Cui TJ 2010
"A class of line transformed cloaks with easily realizable constitutive parameters,"
Jour. Appl. Phys. {\bf 107}, 034911

\end{thebibliography}
\end{document}